\documentclass{aa}

\usepackage[]{graphics}
\usepackage[]{psfig}
\usepackage[]{epsfig}
\usepackage[]{longtable}
\usepackage[]{times}
\include{epsf}
\begin{document}

\title{Discovery of two M32 twins in Abell 1689\thanks{Based on observations obtained at the European
Southern Observatory,
    Chile (Observing Programme 273.B-5008).}}

\author{S. Mieske \inst{1,2} \and L. Infante \inst{2} \and M.
Hilker \inst{1} \and G. Hertling\inst{2} \and J. P. Blakeslee\inst{3} \and N. Ben\'{\i}tez\inst{3} \and H. Ford\inst{3} \and K. Zekser\inst{3}}
\mail{smieske@astro.uni-bonn.de}


\institute{Sternwarte der Universit\"at Bonn, Auf dem H\"ugel 71,
  53121 Bonn, Germany \and Departamento de Astronom\'{\i}a y Astrof\'{\i}sica, Pontificia
Universidad Cat\'olica de Chile, Casilla 306, Santiago 22, Chile \and 
Department of Physics and Astronomy, Johns Hopkins
University, 3400 North Charles Street, Baltimore, MD 21218.}

\date {Received 3 November 2004 / Accepted 9 December 2004}

\titlerunning{M32 twins discovered in A1689}

\authorrunning{S.~Mieske et al.}

\abstract{The M31 satellite galaxy M32 has long been considered an object of unique properties, being the most extreme example of the very rare compact elliptical galaxy class. Here we present the spectroscopic discovery of two M32
twins in the massive galaxy cluster Abell 1689. As these objects are so rare, this is an important step  towards a better understanding of the nature of compact galaxies. The two M32 twins had first been detected within our photometric search for ultra compact dwarf galaxy (UCDs) candidates in A1689 (Mieske et
al.~\cite{Mieske04b}) with the Advanced Camera for Surveys (ACS). Their luminosities ($M_{V}\simeq -17$ mag) are very similar to M32 and their surface brightness profiles are consistent with
that of M32 projected to A1689's distance. From our ACS imaging we detect several fainter compact galaxy candidates with luminosities
intermediate between M32 and the Fornax UCDs. If spectroscopically confirmed as cluster members,
this would almost close the gap in the mag-$\mu$ plane between the region of UCDs and the compact ellipticals, implying a sequence of faint compact galaxies well separated from that of dwarf ellipticals.}

\maketitle

\keywords{galaxies: clusters: individual: Abell 1689 -- galaxies:
dwarf -- galaxies: structure -- galaxies: interactions}

\section{Introduction}
\subsection{Compact ellipticals (cEs) and ultra compact dwarfs (UCDs)}
In the magnitude-surface brightness plane of stellar systems, early-type
galaxies populate a well defined sequence of increasing central surface
brightness with increasing total luminosity (Jerjen \& Binggeli
\cite{Jerjen97a}, Infante et al.~\cite{Infant03}, Hilker et
al.~\cite{Hilker03}, Karick et al.~\cite{Karick03}, Graham \& Guzman~\cite{Graham03}), extending over about ten magnitudes. This sequence is folded up only at the very bright end due to core formation in the most luminous giant ellipticals (Faber et al.~\cite{Faber97}). Naturally, galaxies that fall well outside this sequence are very attractive objects to study. 
Probably the most extreme outlier towards high surface brightness at low luminosity is the compact elliptical (cE) galaxy
M32 (e.g. Faber~\cite{Faber73}, Kormendy~\cite{Kormen85},~\cite{Kormen87}, Nieto \& Prugniel~\cite{Nieto87}). It has $M_V\simeq -17$ mag (Graham~\cite{Graham02}) and an effective surface brightness $\mu_{\rm eff}$ about four mag brighter than dwarf ellipticals (dEs)
of comparable $M_V$. The effective radius of its bulge is only~$\sim$100~pc. 
Systematic searches for M32 analogs in other nearby clusters have up to now
not proven successful (e.g. Ziegler \& Bender~\cite{Ziegle98}, Drinkwater \&
Gregg~\cite{Drinkw98}). There are only few faint elliptical galaxies
with roughly similar properties to M32 (e.g. NGC 4486B and CCC 70, see Jerjen \& Dressler~\cite{Jerjen97b}), but M32 still remains
 the faintest and most compact one.\\
As most of these cEs are found close to more luminous giant galaxies, discussions about their origins have focussed on tidal effects from their neighbours. As progenitor galaxies for M32, bulges of late type spirals (Bekki et al.~\cite{Bekki01}) have been suggested as well as genuine low luminosity ellipticals (Faber~\cite{Faber73}). Choi et al.~(\cite{Choi02}) find that a normal elliptical galaxy cannot have been a precursor of M32, but that its precursor must have been intrinsically compact. Graham~(\cite{Graham02}) finds that M32's surface brightness profile can be best fit by a compact Sersic bulge plus a low surface brightness exponential disk. These two findings support the idea that M32 is the compact bulge of a stripped spiral rather than a normal elliptical.\\
Going about a factor of 100 in luminosity and 10 in effective radius lower, one enters the regime of the so-called ``ultra compact dwarf galaxies" (UCDs), discovered in the Fornax cluster by Hilker
et al. (\cite{Hilker99}) and Drinkwater et al. (\cite{Drinkw00}). These overluminous star clusters have $-11>M_V>-13$ mag (Mieske et al.~\cite{Mieske04a}) and are more concentrated towards the cluster centre than the average dwarf galaxy population. Also for UCDs, tidal stripping of nucleated dwarf galaxies has been proposed as a possible origin (Bekki et al.~\cite{Bekki03}).
\subsection{UCD candidates in Abell 1689}
\begin{figure}
\begin{center}
\epsfig{figure=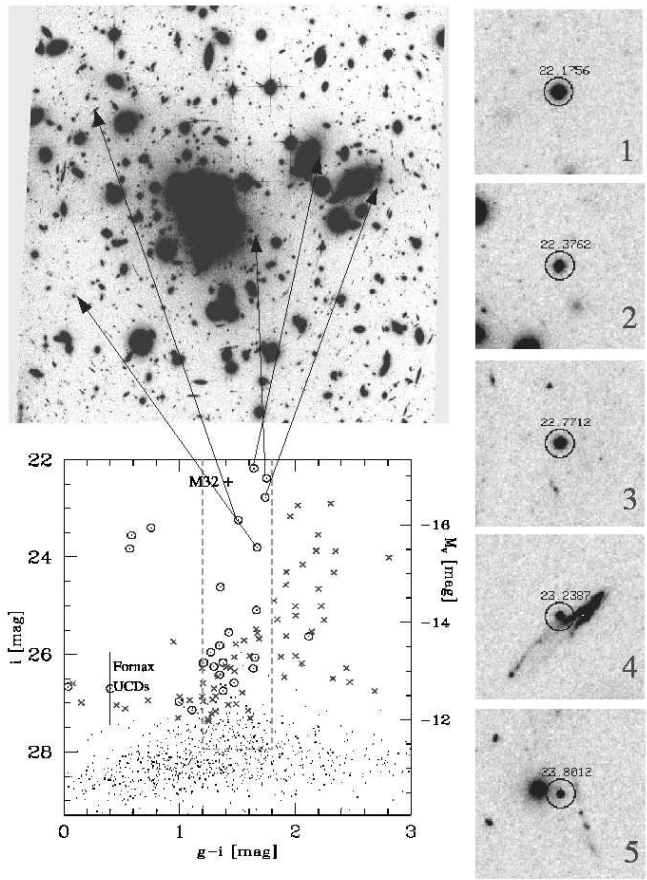,width=8.7cm}
\end{center}
\caption[]{\label{mapcmd}{\it Top:} ACS image of Abell 1689 plus
  thumbnails of the five brightest UCD candidates, numbered by decreasing luminosity according to Table~\ref{resspec}. The thumbnail width is 7$\arcsec$. {\it
    Bottom:} CMD of UCD candidates in Abell 1689. {\it Circles:} Photometric redshift $z_{\rm phot}<0.5$. {\it Crosses:}
 $z_{\rm phot}>0.5$.  The dashed vertical lines indicate the colour range of Fornax UCDs, the vertical tick their apparent luminosities if shifted to A1689's distance (Mieske et al.~\cite{Mieske04a},~\cite{Mieske04b}). The approximate location in the CMD of M32 at A1689's distance is indicated by the cross (Mateo~\cite{Mateo98}).}
\end{figure}

In the densest environments, tidal stripping should occur more frequently than anywhere else.
Therefore we undertook a photometric search for UCD candidates in
the very massive galaxy cluster Abell 1689 ($z=0.183$, $m-M \simeq$ 39.75 mag), see Mieske et
al. (\cite{Mieske04b}). This search was  based on deep high resolution
ACS images, see Fig.~\ref{mapcmd}. UCD candidates were selected based on their compact morphology, brightness and $(g-i)$ colour. In addition, their photometric redshift was demanded to be smaller than 0.5, as derived using the BPZ code by Ben\'{\i}tez (\cite{Benite00}).
The search resulted in the discovery of
several very luminous UCD candidates in the luminosity range between Fornax
UCDs and M32.
Two of
the brightest UCD candidates ($i\simeq$ 22.5 mag, $M_V\simeq -17$ mag) 
are marginally
resolved, implying $r_{\rm eff}\simeq$ 300pc at Abell 1689's distance,
similar to M32.\vspace{0.2cm}\\
In this letter, we present the spectroscopic follow-up of the five brightest UCD candidates in Abell 1689.
\section{Spectroscopic results}
\label{data}
\begin{figure}
\begin{center}
\epsfig{figure=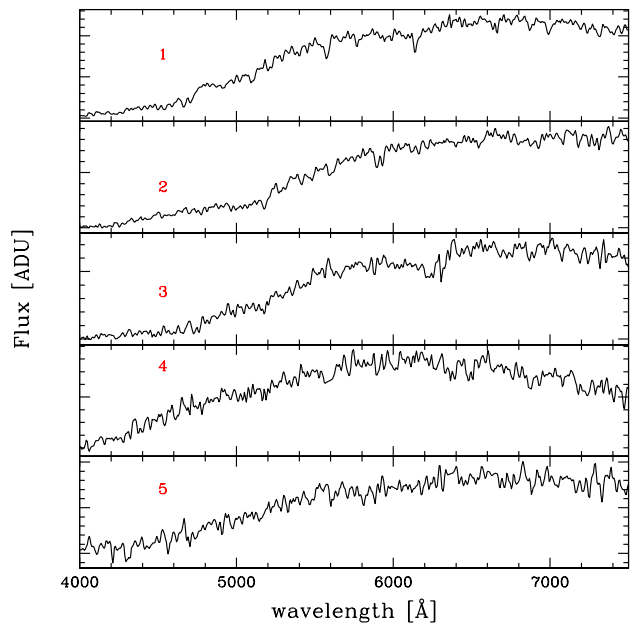,width=8.7cm}
\end{center}
\caption[]{\label{spectra}Wavelength calibrated spectra of the five UCD
  candidates, ordered from top to bottom by decreasing $i$ luminosity. The
  spectra are not flux-calibrated. Candidates 1 and 3 have spectroscopic redshifts consistent with A1689. The H\&K break is at about 4700{\AA} for candidate 1 and 4800{\AA} for candidate 3.}
\end{figure}

The spectroscopic observations have been obtained with
 Director's Discretionary Time (DDT) program 273.B-5008. 
The spectra were taken with the
 VLT/FORS2 mask exchange unit MXU, targeting the five brightest UCD
 candidates in Abell 1689 from Mieske et al.~(\cite{Mieske04b})
and several other objects of interest (stellar super cluster
 candidates, dwarf galaxies, etc.). 
The slit-width was 1$\arcsec$, the resolution was about
 6\AA/pixel at a spatial scale of 0$\farcs 2$ / pixel. The PSF-FWHM was about 0.8$\arcsec$. The total on-source
integration time was 6500 seconds. The
 resulting S/N per pixel in the best exposed wavelength range around 6000
 {\AA} was about 15 for the brightest UCD candidate
 ($i=22.2$ mag) and about 6 for the faintest one ($i=23.8$ mag).
The UCD candidate spectra are shown in Fig.~\ref{spectra}, the derived redshifts in Table~\ref{resspec}.\\
We 
confirm the cluster membership for UCD candidates No.1 and 3. The redshift of candidate No.2 is consistent with a galactic foreground star. The spectra of the two faintest UCD candidates No.4 and 5
were partially blended with neighbouring
galaxies, inhibiting an unambiguous redshift determination. Both spectra exhibit several weak cross correlation peaks in a range of redshifts, including peaks corresponding to A1689's distance. Longer spectroscopic integrations and better seeing are needed to definitely confirm or rule out their cluster membership. Judging from the thumbnail image in Fig.~\ref{mapcmd}, candidate No.4 might also be part of a lensed background galaxy.\\
{\bf A note on nomenclature:} the absolute luminosities of the two confirmed
  A1689 UCD candidates are almost identical to M32, 
such that one might also call them
 ``compact elliptical'' (cE)
  or ``compact dwarf elliptical'' (cdE). We choose to assign the two
  cluster members the more general term ``compact galaxies'' (CGs) 
or ``M32-like galaxies''.
\begin{table*}
\begin{center}
\begin{tabular}{lllllll}
Candidate&Ra[2000]&Dec[2000]&$i$&$(g-i)$&$z_{\rm spec}$ & Comment\\\hline
1&13:11:32.75 & $-$01:19:48.9&22.18 &1.64 &0.1859 $\pm$ 0.0008 &new name: CG$_{\rm A1689,1}$\\
2&13:11:29.90 &$-$01:19:22.5 &22.38 &1.75 &75 $\pm$ 260 km/s&Star \\
3&13:11:33.11 &$-$01:20:05.5 &22.77 &1.74 &0.2014 $\pm$ 0.0007 &new name: CG$_{\rm A1689,2}$\\
4&13:11:31.09 &$-$01:21:14.8 &23.24 &1.51 &$0.00<z_{\rm spec}<0.30$&Possible member (weak, blended)\\
5&13:11:25.73 &$-$01:21:42.1 &23.80 &1.67 &$0.15<z_{\rm spec}<0.20$ &Possible member (weak, blended)\\
\end{tabular}
\end{center}
\caption[]{\label{resspec}Properties of the spectroscopically
  investigated UCD candidates in Abell 1689. ``CG'' in the comment column
stands for ``compact galaxy''. The spectrum of candidate 4 is partially contaminated by a neighbouring irregular galaxy with unkown redshift, see Fig.~\ref{mapcmd}, showing several weak cross-correlation peaks within $0<z<0.30$. The spectrum of candidate is partially contaminated by a dwarf galaxy member of Abell 1689, see Fig.~\ref{mapcmd}, showing several weak cross-correlation peaks within $0.15<z<0.20$.}
\end{table*}
 \section{Comparison between M32 and compact galaxies in Abell 1689}
\label{compm32}
\begin{figure}
\begin{center}
\epsfig{figure=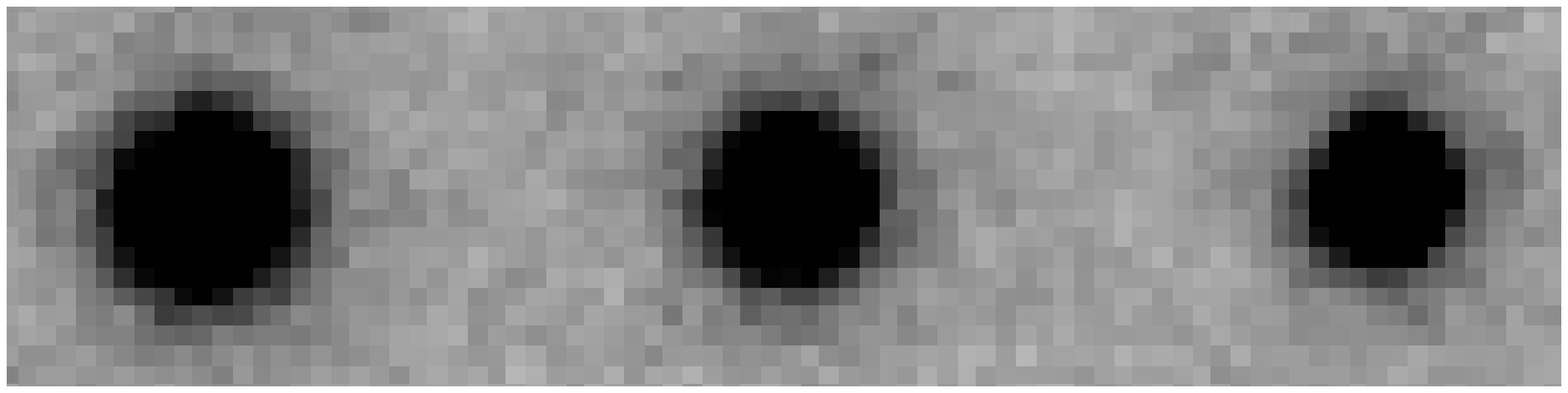,width=8.6cm}
\epsfig{figure=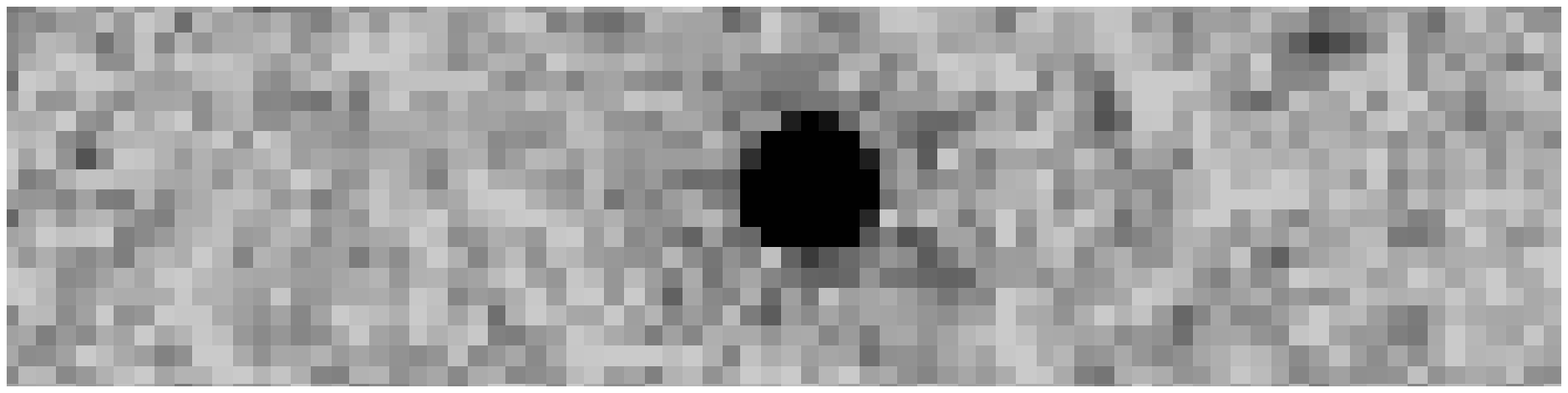,width=8.6cm}
\end{center}
\caption[]{\label{m32comp1}{\it Top panel:} Morphological comparison between
  (from left to right): CG$_{\rm A1689,1}$ ($i=22.2$ mag); M32 modelled
  according to the fit by Graham (\cite{Graham02}), projected to A1689's
  distance and seeing convolved ($i$=22.4 mag); CG$_{\rm A1689,2}$ ($i=22.8$ mag). {\it Bottom panel:}
An unresolved point source ($i=22.4$ mag). The image excerpts have
  the same intensity cuts and are taken
  from deep ACS $i$-band exposures of Abell 1689. There is a higher noise level for
  the lower image because the point source originally is about 1.5 mag
  fainter and was scaled up to fall in the luminosity range of the
  M32-like objects.}
\end{figure}

\begin{figure}
\begin{center}
\epsfig{figure=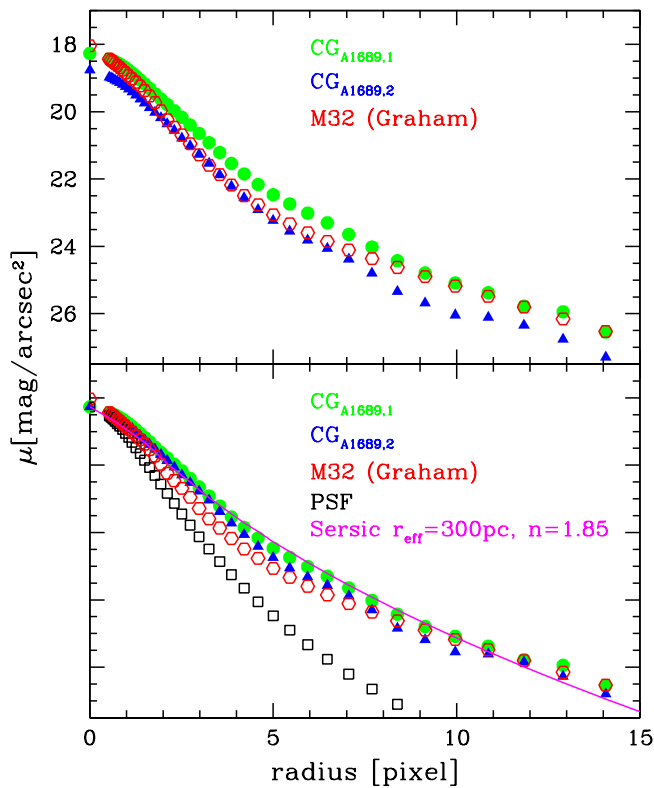,width=8.6cm}
\end{center}\vspace{-0.7cm}
\caption[]{\label{m32comp2}Comparison of surface brightness (SB) profiles, with
  scale
0.05$\arcsec$/pixel, or 155 pc/pixel at Abell 1689's distance. {\it
    Upper panel}: CG$_{\rm A1689,1}$ (filled circles), CG$_{\rm
    A1689,2}$ (filled triangles), M32's profile from the fit by
Graham (\cite{Graham02}),  projected to A1689's  distance and PSF
    convolved (open hexagons). 
{\it Lower panel}: The three SB profiles from the upper
panel normalised to the same peak intensity plus the normalised PSF
    profile (Moffat profile with 0.095$\arcsec$ FWHM). The solid line is a PSF convolved Sersic profile with $r_{\rm eff}=$ 300 pc and $n=1.85$.}
\end{figure}

In Figs.~\ref{m32comp1} and~\ref{m32comp2} the morphological appearance and
surface brightness (SB) profiles of the two A1689 compact galaxies (CGs) 
is compared with a
seeing convolved, simulated SB model of M32, as taken from the
fit by Graham (\cite{Graham02}). This fit consists of a central
bulge component described by a Sersic profile with n=1.5 and $r_{\rm eff}=105$
pc and an underlying exponential disk with exponential scale length $h=480$
pc ($r_{\rm eff}=806$ pc). Ellipticity was adopted as zero. The PSF profile used for convolution
was adopted as a Moffat function with FHWM$=0.095''$ and $\beta=2.5$.
For the simulation, an angular distance scale of
3.10 kpc/$\arcsec$ was assumed, corresponding to a mean cluster redshift of
$z=0.1832$ (NED database), $H_{\rm 0}=70$ km s$^{-1}$ Mpc$^{-1}$, $\Omega_M=$
0.3 
and $\Omega_{\Lambda}=0.7$.\\
On the
ACS image in Fig.~\ref{m32comp1}, the
two A1689 CGs and the simulated M32 appear very similar and are notably more extended
than a pure stellar source of comparable luminosity. 
In the surface brightness plots of Fig.~\ref{m32comp2}, this is confirmed. When fitting an exponential profile
to the outer parts of the two CG profiles that are mostly unaffected by the seeing, effective radii in the range
700-800 pc are determined, in agreement with the scale size of M32's exponential
disk. In the lower panel of
Fig.~\ref{m32comp2}, it is shown that also a single Sersic profile with
$r_{\rm eff}=300$ pc and $n=1.85$ fits
quite well the observed CG profiles. 
In order to explore the acceptable
range for $r_{\rm eff}$ in the case of a single component, we used the ISHAPE
task (Larsen~\cite{Larsen99}). 
As input PSFs we used three different unresolved sources in the $i$
exposure of the ACS image, with PSF-FWHM between 0.09 and 0.10$''$. For
the intrinsic profile shape we adopted
King profiles with concentration parameter between 5 and 30 and a Moffat profile
with $\beta= 2.5$. The mean and scatter of the fitted $r_{\rm eff}$ obtained
with different models and PSFs is 370 pc $\pm$ 180 pc for CG$_{\rm A1689,1}$
and
225 pc $\pm$ 75 pc for CG$_{\rm A1689,2}$. All of the profile fits have a
comparable $\chi ^2$. Doing the same procedure for the simulated, seeing
convolved M32 image yields an intrinsic effective radius of 225 pc $\pm$ 75
pc.
\label{M32comp}
\section{Discussion and conclusions}
\label{discussion}
\begin{figure}[h!]
\begin{center}\vspace{-1cm}
\epsfig{figure=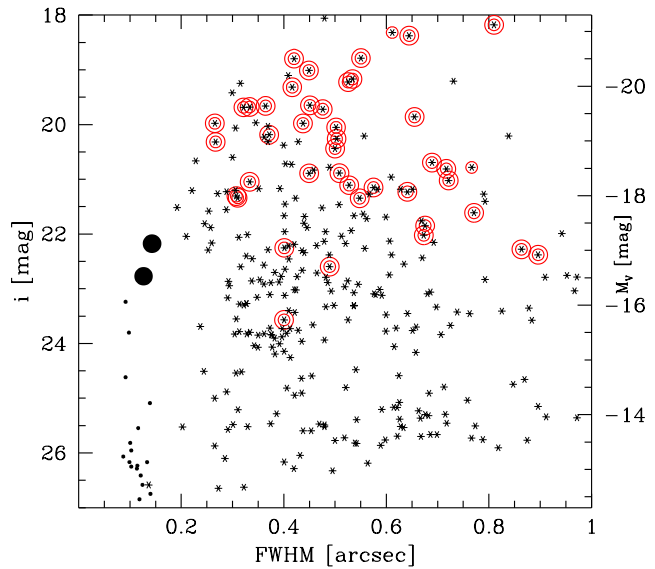,width=8.6cm}\vspace{-0.5cm}
\end{center}
\caption[]{\label{sizelum}Size-luminosity distribution of normal dwarf galaxies and UCD candidates in Abell 1689. SExtractor FWHM is plotted vs. $i$-band luminosity for sources with $z_{\rm phot}<$ 0.5 in the indicated magnitude range. The PSF-FWHM is about 0.10$\arcsec$. The dots are the unresolved UCD candidates, defined by a star-classifier larger than 0.6 (see Mieske et al.~\cite{Mieske04b}). The asterisks indicate the resolved dwarf galaxy candidates, defined as having SExtractor star-classifier smaller than 0.6. The larger filled circles indicate the two confirmed M32 twins. The open circles indicate galaxies with a measured spectroscopic redshift (Czoske~\cite{Czoske04}), double circles are those with $0.16<z<0.22$ (redshift range of Abell 1689).}
\end{figure}
Objects like M32 are extremely rare. In this letter we have presented the discovery of two M32 analogs in Abell 1689, an extremely massive cluster which is also known as the strongest lens on the sky ($M \simeq 1 \times 10^{15} M_{\sun}$, see King et al.~\cite{King02}). This is consistent with the hypothesis that deep potential wells favour the creation of such compact galaxies. CG$_{\rm A1689,1}$ has a relative velocity of $-$4800 km s$^{-1}$ with respect to the closest projected cluster giant elliptical and $+$1500 km s$^{-1}$ to the second nearest one, values which are consistent with the high velocity dispersion of Abell 1689 (Czoske~\cite{Czoske04}). CG$_{\rm A1689,2}$ has a much smaller velocity difference of about 100 km s$^{-1}$ to its closest neighbour, comparable to the relative velocity between M31 and M32. 
This fits into a picture where both galaxy-satellite interactions and tidal interaction with the overall cluster potential can lead to the creation of compact galaxies.
If this is true then there should exist a continuous sequence of compact galaxies between the fainter Fornax-UCD types and brighter M32-types, especially in an extremely massive cluster like Abell 1689. As is shown in Figs.~\ref{mapcmd} and~\ref{sizelum} and outlined in detail in Mieske et al.~(\cite{Mieske04b}), there indeed is a population of compact galaxy {\it candidates} with luminosities intermediate between the Fornax UCDs and M32, in addition to the two brighter confirmed M32 analogs. The definite confirmation of their membership with spectroscopy is needed to exclude them being foreground stars or background galaxies. If confirmed as cluster members, this would fill a previously unpopulated region in the mag-$\mu$ plane of stellar systems: a sequence of faint compact galaxies parallel to the dwarf galaxy sequence but offset towards higher surface brightness / smaller scale length.  Fig.~\ref{sizelum} illustrates this: there is a notable gap in FWHM between the unresolved UCD candidates and the resolved ``normal'' dwarf galaxy candidates. Our sample is complete across this gap, i.e. the gap is not caused by selection effects.\\
The next step will be to perform a spectroscopic census of the entire
compact galaxy population in Abell 1689, extending from the brighter cE
candidate galaxies not observed by Czoske~(\cite{Czoske04}) to several mag fainter than the two newly discovered M32 twins. Furthermore, the results presented here encourage us to propose a search for compact galaxy candidates in other
massive clusters.
\acknowledgements
\noindent We thank the ESO User Support Group for carrying out the DDT
spectroscopy (program 273.B-5008) in service mode. ACS was developed under
NASA contract NAS 5-32864. SM was supported by DAAD Ph.D. grant Kennziffer D/01/35298 and
DFG Projekt Nr. HI 855/1-1. LI would like to acknowledge support from ``proyecto Fondap \# 15010003".

\end{document}